# Effect of neutron irradiation on the properties of FeSe compound in superconducting and normal states.

A. E. Karikin[1], T. Wolf[2], A. N. Vasil'ev[3], O. S. Volkova[3], and B. N. Goshchitskii[1].

[1] *Institute of Metal Physics, Ural Branch, RAS, 620219 Ekaterinburg, ul. S. Kovalevskoi 18*
[2] *Institut für Festkörperphysik, D-76021 Karlsruhe, Germany.*
[3] *Moscow State University, 119992 Moscow, Vorob'evy Gory*

e-mail: aekarkin@rambler.ru

Effect of atomic disordering induced by irradiation with fast neutrons on the properties of the normal and superconducting states of polycrystalline samples FeSe has been studied. The irradiation with fast neutrons of fluencies up to $1.25 \cdot 10^{20}$ cm$^{-2}$ at the irradiation temperature $T_{irr} \approx 50$ °C results in relatively small changes in the temperature of the superconducting transition $T_c$ and electrical resistivity $\rho_{25}$. Such a behavior is considered to be traceable to rather low, with respect to that possible at a given irradiation temperature, concentration of radiation defects, which is caused by a simpler crystal structure, considered to other layered compounds.

The discovery of high-temperature superconductivity in layered Fe-based compounds [1] gave rise to experimental and theoretical researches into these systems. A systematic study of the disordering effects in new superconductors is especially important in the case of possible anomalous types of the Cooper pairing (anisotropic, *s*, *p,* or *d*-pairing, triplet pairing, etc.) [2]. Investigating the influence of the fast-neutron irradiation on the electronic, magnetic, and other properties of materials, one can gain new information on the properties of an pristine (ordered) system, which can be exemplified by a qualitatively different behavior of superconductive properties under irradiation of the $MgB_2$ [3, 4, 5] and $YBa_2Cu_3O_7$ [3, 6, 7] superconductors. The former, relatively weak decrease of $T_c$, which is typical of the systems with strong electron-phonon interaction, and the latter, rapid and complete, degradation of superconductivity, explicitly purports another more exotic mechanism of pairing. So, studying response of a superconductive system to the neutron irradiation allows one to gain additional data on the nature of superconductivity: mechanism of pairing, symmetry of the superconductive gap, etc. Earlier studies of the disordering effects induced by irradiation with fast neutrons in the compound La(O-F)FeAs showed a rapid degradation of superconductivity, which points to an anomalous type of the Cooper pairing (most likely, anomalous isotropic $s^{\pm}$ type) [8].

In the given work we present the results of study of the influence of atomic disordering on the properties of the normal and superconductive states of the polycrystalline samples of the FeSe system, which possesses the simplest crystal structure among the recently discovered Fe-based superconducting compounds. If in other compounds of this type the Fe-As layers alternate with the layers of the type La-O (system 1111), Li (111), Ba (122), and to realize superconductivity, it is necessary to dope the compounds with elements of different valence, the Fe Se structure does not contain either parceling layers or doping elements. The value of $T_c \approx 10\text{-}15$ K in FeSe is somewhat lower than in other compounds of this class $T_c \leq 55$ K [9], but it is increased in FeSe to $\approx 30$ K via applying external pressure [10] or intercalating with alkaline elements [11], and this presents grounds for a supposition on a common mechanism of superconductivity in all these compounds. However, the rather simple crystal structure of FeSe causes, as will be shown in what follows, poor susceptibility of the system to radiation defects which are formed upon the fast-neutron irradiation.

The polycrystalline FeSe$_{0.963}$ samples were obtained by conventional solid-state synthesis [12]. The electrical conductivity $\rho$ and Hall coefficient $R_H$ were measured on polycrystalline samples with dimensions $2 \times 2 \times 0.2$ mm$^3$ using a standard 4-probe method [13] in the temperature range $T = 1.5 - 380$ K and at magnetic fields of up to 13.6 T. Three samples cut from one and the same ingot were irradiated with fast neutrons at the temperature $T_{irr} = (50\pm10)$°C: sample №1 with fluencies 0.5, 1.5, 2.5, 4.5, and $12.5 \cdot 10^{19}$ cm$^{-2}$; sample №2, 1.0 и $2.0 \cdot 10^{19}$ cm$^{-2}$; sample №3, $8.0 \cdot 10^{19}$ cm$^{-2}$. The pristine samples had close values of $T_c \approx 10$ K and similar temperature dependences of the electrical resistivity $\rho(T)$ and Hall coefficient $R_H(T)$.

The value of $R_H(T)$ is positive in the region of low temperatures $T < 100$ K, negative at $100$ K $\leq T < 200$ K, and again changes its sign at $T \geq 200$ K, which reflects a multiband electronic structure of this compound. The characteristics $\rho(T)$ is power-dependent at low temperatures; at $T \geq 200$ K this dependence flat-tens, and at $T \approx 350$ K the slope $d\rho/dT < 0$, which can be related to the appearance of a contribution to conductivity of the activation type. (Fig. 1).

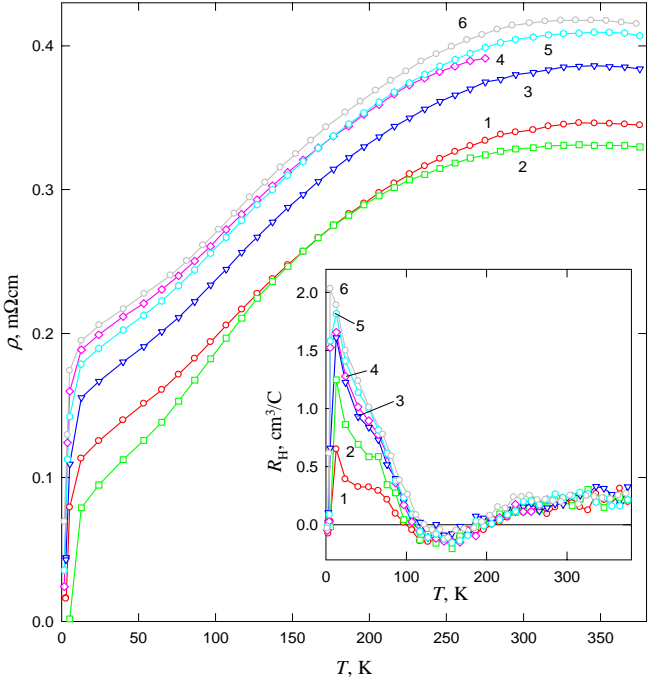

Fig. 1. Temperature dependences of the electrical resistivity $\rho$ for the FeSe sample №1: in the pristine (1) and irradiated states at fluencies of fast neutrons $0.5 \cdot 10^{19}$ cm$^{-2}$ (2), $1.5 \cdot 10^{19}$ cm$^{-2}$ (3), $2.5 \cdot 10^{19}$ cm$^{-2}$ (4), $4.5 \cdot 10^{19}$ cm$^{-2}$ (5), and $12.5 \cdot 10^{19}$ cm$^{-2}$ (6). The inset shows temperature dependences of the Hall coefficient $R_H$ in the magnetic field $H = 13.6$ T.

The irradiation does not cause any qualitative changes in the temperature dependences $\rho(T)$ and $R_H(T)$. Upon relatively low fluencies $\Phi \leq 1.0 \cdot 10^{19}$ cm$^{-2}$ there is observed a decrease of $\rho$ at low temperatures (Fig. 1) and a slight increase of $T_c$ (Fig. 2). This effect is accompanied by a remarkable increase of $R_H$ at low temperatures, whereas in the high-temperature range the changes are insignificant. At $\Phi > 1.0 \cdot 10^{19}$ cm$^{-2}$ there takes place a decrease of $T_c$, growth of $\rho(T)$, mainly due to growing residual resistivity, and a further increase of $R_H(T)$ at low temperatures (Fig. 1). Fig. 2 shows the behavior of several parameters depending on the fluence of fast neutrons $\Phi$: temperature of superconducting transition $T_c$, which was taken in the middle of the resistivity transition; electrical resistivity $\rho_{25}$ measured at $T = 25$ K; and the Hall coefficient $R_{Hmax}$ (maximal value in the low-temperature region). These parameters virtually cease to change at $\Phi > 2.0 \cdot 10^{19}$ cm$^{-2}$, which means that an ultimate value of the concentration of radiation defects, which is an equilibrium one at the given temperature $T_{irr} \approx 50$ °C, is achieved. A relatively small increase of the residual resistivity 0.1–0.2 mΩ·cm upon irradiation indicates that this ultimate concentration is very low. To compare, under irradiation of La(O-F)FeAs with fast neutrons of the fluence $1.6 \cdot 10^{19}$ cm$^{-2}$, which results in the complete degradation of superconductivity, the low-temperature value of $\rho_{40}$ increases more than by an order of magnitude: from $\approx 0.2$ to $\approx 4$ mΩ·cm [8].

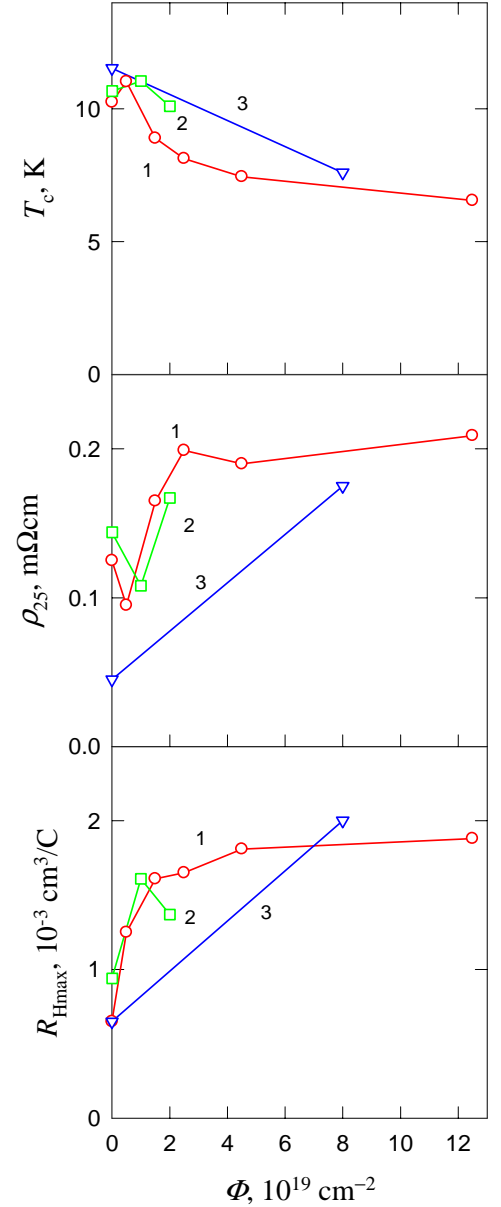

Fig. 2. Temperature of superconducting transition $T_c$ taken in the middle of the transition, electrical resistivity at $T = 25$ K $\rho_{25}$, and Hall coefficient $R_{Hmax}$ depending of the fast-neutron fluence for the FeSe samples: №1 (1), №2 (2), and №3 (3).

Yet, the rate of decreasing $T_c$ as a function of $\rho_{25}$ (Fig. 3), turns out quite comparable with that observed upon irradiation of La(O-F)FeAs [8]. The linear extrapolation gives the complete suppression of superconductivity at $\rho_{25} \approx 1$ mΩ·cm, compared to $\rho_{40} \approx 4$ mΩ·cm in La(O-F)FeAs. Thus, radiation defects also serve to effectively suppress $T_c$ in FeSe, just as in La(O-F)FeAs, which testifies to an anomalous type of the Cooper pairing in FeSe. However, the crystal lat-tice of FeSe turns out far more stable with respect to the formation of defects. Such radiation resistance of FeSe can be related to different, in comparison with La(O-F)FeAs, possible types of defects, instability of the substitution defects Fe − Se caused by the essential difference in the atomic volumes of Fe and Se (11.8 and 27.2 Å$^3$, respectively), etc.

In the inset in Fig. 3 the slope of the second critical field $dH_{c2}/dT$-vs.- electrical resistivity $\rho_{25}$ function is shown. This relatively weak, roughly linear, dependence shows that both the pristine and irradiated samples correspond to the region of "dirty" superconductor for which the electron free path $l$ is on the order of coherency length $\xi$, which is also a sequence of the relatively poor change in the defect concentration upon irradiation.

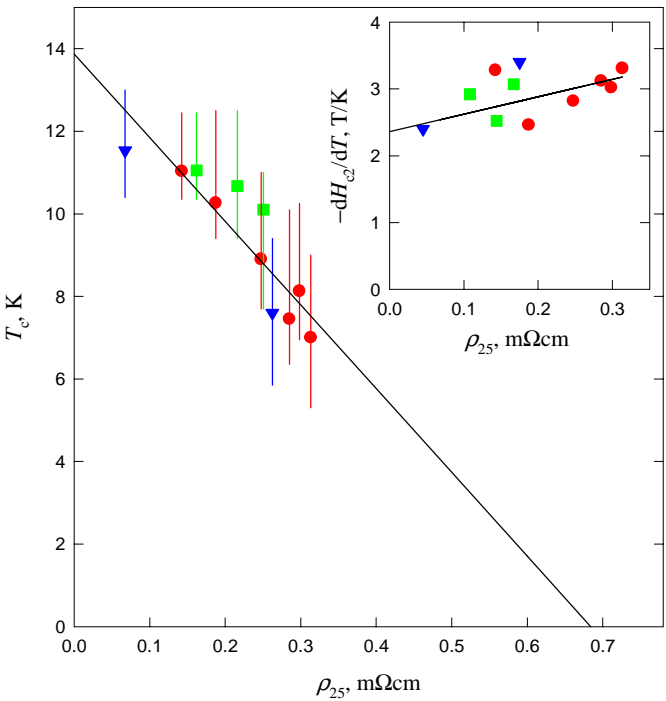

Fig. 3. $T_c$ and $dH_{c2}/dT$ (inset) depending on electrical resistivity $\rho_{25}$ for the FeSe samples: №1 (●), №2 (■), №3 (▼). Vertical lines mark the width of transition into superconducting state at the points of 10% and 90% the normal electrical resistivity; solid lines show linear extrapolation.

The effect of low fluencies at which there are observed a decrease of $\rho$ and increase of $T_c$ (Fig. 2) is unlikely to originate from certain processes that enlarge the degree of ordering. The significant growth of $R_{Hmax}$ in this region points to the effects of doping. In the given case this is the increase in the concentration of the hole-type carriers due to the positive effective electrical charge of the radiation defects, which results in a shift of the Fermi level toward the low energies. The observed increase in the concentration of hole possessing high mobility in the region of low temperatures gives a small (on the order of 0.05 mΩ·cm) decrease of $\rho_{25}$ and, besides, increase of $T_c$ by ~0.5-0.8 K.

The work was supported by the Plan of RAS (theme № 01.2.006 13394, "PULSE") and partially by the Program for Fundamental Research of the Presidium of RAS "Quantum Physics" (grant № 09-П-2-1005 Ural Branch, RAS), the Russian Foundation for Basic Research, (grant № 11-02-00224), the Ministry for Science and Education (State Contract № 16.518.11.7032).